\RequirePackage{snapshot}
\documentclass[conference]{IEEEtran}

\usepackage[symbol]{footmisc}

\usepackage{color,latexsym,amsfonts,amssymb}
\usepackage{hyperref}
\usepackage{amsmath,cite}
\usepackage{amsmath,amsthm,amsbsy}
\usepackage[mathscr]{euscript}
\usepackage[pdftex]{graphicx} 
\usepackage{tikz} 
\usetikzlibrary{math} 
\usetikzlibrary{calc}
\usepackage{pgfplots} 
\pgfplotsset{compat=1.18} 
\usepackage{bbm}
\usepackage{authblk}

\usepackage{mathtools} 
\usepackage{caption} 
\usepackage[normalem]{ulem}

\newcommand{\blue}[1]{\textcolor{blue}{#1}}

\newcommand{\cyan}[1]{\textcolor{black}{#1}}

\newcommand{\R}{\mathbb{R}}

\newcommand{\eqdef}{=\vcentcolon}

\newcommand{\SINR}{\textnormal{SINR}}

\newcommand{\noise}{N}
\newcommand{\fading}{F}
\newcommand{\power}{P}

\newcommand{\scale}{\kappa}

\newcommand{\User}{X^{\lambda}}

\usepackage[compact]{titlesec}
  \titlespacing{\section}{0pt}{0.5ex}{.5ex}
   \titlespacing{\subsection}{0pt}{0.5ex}{.5ex}
   \titlespacing{\subsubsection}{0pt}{0.5ex}{0.5ex}

\theoremstyle{plain} 

\theoremstyle{definition}

\title{Connectivity and interference in  device-to-device networks in Poisson-Voronoi cities}

\author[1]{H. P. Keeler~\thanks{h.keeler@unimelb.edu.au}}
\author[2]{B. B{\l}aszczyszyn~\thanks{Bartek.Blaszczyszyn@ens.f}}
\author[3]{E. Cali~\thanks{elie.cali@orange.com }}

\affil[1]{University of Melbourne, Australia}
\affil[2]{Inria/ENS, Paris, France}
\affil[3]{Orange Labs,  Paris, France}

\begin{document}



\maketitle

\begin{abstract}
To  study the overall connectivity in  device-to-device networks in cities, we incorporate a signal-to-interference-plus-noise connectivity model into a Poisson-Voronoi tessellation model representing the streets of a city. Relays are located at crossroads (or street intersections), whereas (user) devices are scattered along streets. Between any two adjacent relays, we assume data can be transmitted either directly between the relays or through users, given they share a common street. Our simulation results reveal  that the network connectivity is ensured when the density of users (on the streets) exceeds a certain critical value. But then the network  connectivity disappears when the user density exceeds  a second critical value. The intuition is that for longer streets, where direct relay-to-relay communication is not possible,  users are needed to transmit data between relays, but with too many users  the interference becomes too strong, eventually reducing the overall network connectivity. This observation on the user density evokes  previous results based on another wireless network model, where transmitter-receivers were scattered across the plane. This effect disappears when interference is removed from the model, giving a variation of the classic Gilbert model and recalling the lesson that neglecting interference in such network models can give overly optimistic results. For physically reasonable model parameters, 
we show that crowded streets (with more than six users on a typical street) lead to a sudden drop in  connectivity. We also give numerical results outlining a relationship between the user density and the strength of any  interference reduction techniques. 
\end{abstract}

\section{Introduction}
\footnote[0]{Bartłomiej Błaszczyszyn’s
work was partly supported by the European Research Council (ERC NEMO-788851).}
Imagine a city with phone relay stations, located at various crossroads (or street intersections), and, scattered along each street are people willing to relay data through their devices, forming a large device-to-device network. Can we relay data through such a network? What happens when the number of people on the streets increases? How does the network connectivity behave?  Motivated by recent mathematical results~\cite{bartek2023sir},  we will present supporting  numerical results that highlight how increasing the user (or device)  density   can increase the overall (or macroscopic) network connectivity, but too many users will eventually destroy the connectivity due to interference. We place a focus on  network connectivity and the ability to reduce interference in the network.

\subsection{Percolation theory} 
For large networks, the above types of questions motivate the field of probability known as \emph{percolation theory}, which has its historical origins in studying physical materials and wireless networks~\cite{grimmett1999percolation,meester1996continuum}. Percolation theory studies systems of infinitely large sets of elements, such as particles and transmitter-receiver pairs, and how the system behaves when connections (or bonds) are randomly formed between the elements, creating so-called \emph{clusters} or \emph{components}. When a connection exists between two elements, we say the connection is \emph{open}, otherwise it is \emph{closed}. When an infinitely large component arises, we say that \emph{percolation} has occurred or the system \emph{percolates}. The mechanism through which the random connections form depends on model parameters. For a single parameter,  its value when percolation occurs is called the \emph{critical value} or \emph{percolation threshold}, while for multiple parameters, we use the term \emph{percolation regime}. When there is no percolation, we say the system is in a \emph{subcritical} state. Conversely, when the system percolates, it is in a \emph{supercritical} state, which, for wireless network models,  means that in theory connectivity is assured across the network. Researchers have used the tools from percolation theory to study and gain insight into wireless networks~\cite{gilbert1961random,dousse2005impact,dousse2006percolation,franceschetti2007random}.       

\subsection{Poisson-Voronoi tessellations as city street systems} 
A \emph{Voronoi} (or \emph{Dirichlet}) \emph{tessellation} is a fundamental geometric object with countless applications. This tessellation constructed from a Poisson point processes is highly studied due to it being a tractable yet insightful model~\cite{moller1994poisson,okabe2000spatial}. Using it,  
Le Gall, B{\l}aszczyszyn, Cali and En-Najjary~\cite{le2021continuum} recently presented a new percolation model to study the communication abilities of wireless device-to-device networks with relays located at crossroads. They 
assumed the edges of all the Voronoi cells represent streets in a large city,  which is an assumption supported by empirical work~\cite{costestim,fittingmontecarlo}.
On each street, they assumed that the users are scattered according to an independent Poisson point process. This percolation model uses a \emph{line-of-sight} requirement, where relays and users (or devices) can only communicate with each other when they share a street. The model also uses a connectivity requirement that is purely geometric, where users and relays can only communicate with each other if they are within some fixed distance of each other. This simple idea for wireless communication builds off a classic model in (continuum) percolation theory originating from the pioneering work by Gilbert~\cite{gilbert1961random}. 

The current paper is motivated by recent mathematical results in a yet-to-be-published paper~\cite{bartek2023sir} in which we enlarge the above device-to-device percolation model~\cite{le2021continuum} by introducing a connectivity model based on the \emph{signal-to-interference-plus-noise ratio}. This fundamental concept gives an upper bound on the successful communication, hence forming the basis of most recent wireless network models~\cite{dousse2005impact,dousse2006percolation,franceschetti2007random}. 

\subsection{Contributions} 
In recent work~\cite{bartek2023sir} we show 
mathematically that percolation is possible in our network model. But this work does not include any numerical work nor indicate the critical parameter regime for percolation. The current work fills that gap by reporting on the results of simulating and studying our network model. 
 
More specifically, for a device-to-device network, we examine the connectivity effects induced by increasing the intensity or (average) density of users $\lambda$. Provided small enough transmission ranges between relays, due to noise and signal attenuation (or path loss),  we numerically demonstrate that the network does not percolate for low user densities, as there are not enough users on the streets.  But as we increase the user density $\lambda$, we observe that the network percolates when the density reaches a  certain critical value $\lambda_1^*$.  If we continue to increase the user density beyond this first critical value, we observe that the network eventually stops percolating as the user density exceeds a second critical value $\lambda_2^*$. Put another way, our simulation results strongly suggest that the  density $\lambda$
 for connectivity is located in a parameter window such that $0\leq \lambda_1^* \leq \lambda \leq \lambda_2^* < \infty$.

 We also study the network model for a range of interference reduction values. We highlight the sharp sensitivity of the critical values with respect to the model parameters such as length scale and noise. Using physically reasonable parameter values, we show for which user densities a network with such parameters stops percolating.

\subsection{Related work}
\subsubsection{Percolation models}
To study wireless networks, Gilbert created the field of continuum percolation with his pioneering paper~\cite{gilbert1961random} by introducing a novel random spatial model. It used a (homogeneous) Poisson point process for the locations of transmitters and receivers. 
The Poisson case has now been extensively studied; see the monograph by Meester and Roy~\cite{meester1996continuum}. 
Using a (homogeneous) Poisson point process, Dousse, Baccelli, and Thiran~\cite{dousse2005impact} did the first work to incorporate interference into a percolation model, which was developed further mathematically in another paper~\cite{dousse2006percolation}. These seminal papers~\cite{dousse2005impact,dousse2006percolation} demonstrated the balance between interference reduction and network density. Meester and Franceschetti~\cite{franceschetti2007random} detailed this line of work in their monograph, which gives a good introduction to signal-to-interference percolation. 

We present here a continuum percolation model that is based on a Cox point process, which effectively adds a layer of randomness to the standard Poisson model by making its density (measure) also random. To study this type of percolation, Hirsch, Jahnel and Cali~\cite{hirsch2019continuum} introduced a general framework requiring two properties in a Cox percolation model. (The first property requires a degree of (decaying) spatial dependence, whereas the second property requires a degree of connectedness.) 
T{\'o}bi{\'a}s~\cite{tobias2019message,tobias2020signal} studied  percolation of a Cox model based on the signal-to-interference. T{\'o}bi{\'a}s and Jahnel~\cite{jahnel2022sinr} studied signal-to-interference percolation for Cox point processes with random (signal) powers. They showed mathematically that percolation was possible in their model provided certain moment conditions on signal powers. 

Le Gall, B{\l}aszczyszyn, Cali and En-Najjary wrote papers on device-to-device networks~\cite{le2021continuum,le2019influence,le2020relay} and used the Cox framework to prove percolation results for a zero-interference (geometric) Gilbert-type model coupled with the line-of-sight assumption~\cite{le2021continuum}. In addition to the mathematical results, Le Gall, B{\l}aszczyszyn, Cali and En-Najjary  used simulations to study percolation in their Gilbert model, producing a paper ~\cite{le2019influence}, which shares a similar spirit to our current paper. 

\subsubsection{Poisson-Voronoi tessellations}
There is an abundance of literature on the Poisson-Voronoi tessellation~\cite{moller1994poisson,okabe2000spatial}.
There are efficient algorithms  for simulating a single Voronoi cell to study its properties. Although not suitable for our purposes, since the cells of our tessellation influence one another, this approach gives a fast way to generate a single Voronoi cell; see M{\o}ller\cite[Section 4.58]{moller1994poisson}. There are also fast simulation methods for generating Poisson-Voronoi tesselations on a torus, giving tessellations with so-called \emph{periodic boundary conditions}; see the method comparison by Su and Drysdale~\cite{su1995comparison}. 
But, again, these approaches are not suitable for our purposes because they introduce a distance distortion when mapping the plane to a torus, thus they would introduce a bias into our model. We use a more straight forward simulation method, which we briefly describe in Section~\ref{sec.simulation}.  Gafur et al.~\cite{Gafur2018} observed that this method can be as fast and accurate for studying percolation thresholds.

\subsubsection{Simulating percolation models}
 Newman and Ziff  wrote an influential paper~\cite{newman2001fast} in which they presented a fast method for simulating discrete percolation systems. They demonstrated  their method by producing numerical results for a square lattice and a spin-glass model in physics. Mertens and Moore~\cite{mertens2012continuum} adapted this approach to the relevant case of continuum percolation for the Gilbert model based on a Poisson point process. Observing that percolation components are simply unions, a principal part of the Newman-Ziff approach relies upon using (fast) union-find algorithms, which have been studied extensively~\cite{tarjan1984worst}
 
 Unfortunately, union-find algorithms assume monotonic connectivity (adding new objects never breaks up existing sets of objects), whereas signal-to-interference models lack this complexity-reducing property because adding a user can both break or create links; discussed further in Section~\ref{sec.nonmonotonic}.  Instead of finding unions of objects and tracking disjoints sets, we need to track disjoint sets that  can merge or break when adding new objects. This a more recent research topic known as \emph{dynamic connectivity}~\cite{henzinger1999randomized,thorup2000near,holm2001poly}, which we consider too involved for our current purposes. (There is also a dearth of numerical libraries using these methods.) But perhaps there is  potential in simplifying these approaches when factoring in the underlying geometry of  signal-to-interference-plus-noise models.

\section{Model}
\subsection{Communication}
\subsubsection*{Signal-to-interference-plus-noise}
We consider a finite collection of transmitter-and-receivers $\{z_1,\dots, z_n\}$ located in the plane $\R^2$.  We denote $P_{i,j}$ as the power of the signal received at $z_j$ originating from a transmitter at $z_i$. We define the  \emph{signal-to-interference-plus-noise ratio (SINR)}  at $z_j$ with respect to the incoming signal from a transmitter at $z_i$ as 
\begin{equation}\label{eq.def.sinr}
\SINR(z_i,z_j) \eqdef  \frac{P_{i,j}} {\noise +\theta I_{i,j} } \,, \quad z_i\neq z_j \,,
\end{equation}
where $\noise\geq0$ is just a noise constant, $P_{i,j}$ is the power of a signal received at $z_j$ originating from a transmitter at $z_i$, and $\theta$ is a technology-dependent parameter, which we can call the \emph{interference reduction parameter}, and  
\begin{equation}
 I_{i,j}=\sum\limits_{k\neq i, j} P_{k,j} 
\end{equation}
is the \emph{interference}, meaning the sum of signals from transmitters $\{ z_1,\dots, z_n \}\setminus\{z_i,z_j\}$. For brevity, we sometimes write just \emph{signal-to-interference}, but we are still assuming a non-zero noise term $\noise>0$.

\subsubsection*{Path loss }
Consider a transmitter at $z_i\in \R^2$ and receiver at $z_j\in \R^2$. The standard path loss model assumes the signal power $P_{i,j}$ takes the form $P_{i,j}= \power_i\ell \left(\Vert z_i-z_j \Vert \right) $.
Here $\power_i$ is the power of the transmitted signal at the source $z_i$. We assume that all transmitters share a fixed power $\power$, meaning $P_i=P$ for all $z_i$. Then we see in the  signal-to-interference model that we can replace $P$ with unit power and replace $N$ with $\bar{\noise}\eqdef\noise/P$.
(Often propagation models also include an additional random variable $\fading_i$ representing propagation phenomena such as fading, but we do not  include it in this work.)

We assume the following properties for our path loss model $\ell(d)$, where the distance $d\geq0$.
\begin{enumerate}
\item $\ell(d)$ is a non-negative function 

\item $\ell(d)$ is a continuous, decreasing path loss function in $r$.

\end{enumerate}
The above properties are physically intuitive. Additionally, Dousse et al.~\cite{dousse2006percolation} showed for their signal-to-interference percolation model that the property below is needed for proving that the network percolates. 
\begin{enumerate}  \setcounter{enumi}{2}
\item  $\ell(0)> \tau\noise/\power$.

\end{enumerate}

\subsection{Network model}

\subsubsection*{Street system $S$}
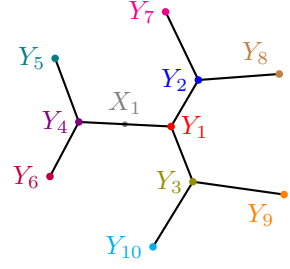
\begin{figure}
\centering
\begin{tikzpicture}[thick,scale=0.3, every node/.style={scale=1}]
\tikzmath{
\x1 =1.9836; \y1 =-0.2124; \r1 =3.1072;
\x2 =3.1904; \y2 =1.8493; \r2 =3.041;
\x3 =2.9483; \y3 =-2.6616; \r3 =3.0041;
\x4 =-2.1058; \y4 =-0.010985; \r4 =3.2709;
\x5 =-3.1565; \y5 =2.809; \r5 =3.3782;
\x6 =-3.3861; \y6 =-2.4297; \r6 =3.3346;
\x7 = 1.7380 ; \y7 =4.8734;\r7 =2.9859;
\x8 =6.7752; \y8 =2.1163; \r8= 3.1533 ;
\x9 =6.9899 ; \y9 =-3.2304; \r9 =3.3043;
\x0 =1.1798 ; \y0 =-5.5495; \r0 =3.2357;
\xp1 =-5.468; \yp1 =-5.0345;
\xp2 =-5.3659; \yp2 =0.25354;
\xp3 =-5.5607; \yp3 =5.1822;
\xp4 =-0.053958; \yp4 =-2.5582;
\xp5 =0.18643; \yp5 =2.3223;
\xp6 =4.405; \yp6 =-5.2888;
\xp7 =5.0736; \yp7 =-0.53846;
\xp8 =4.6991; \yp8 =4.4897;
}

\coordinate (V1) at (\x1,\y1);
\coordinate (V2) at (\x2,\y2);
\coordinate (V3) at (\x3,\y3) ;
\coordinate (V4) at (\x4	,\y4) ;
\coordinate (V5) at (\x5,\y5) ;
\coordinate (V6) at (\x6,\y6) ;
\coordinate (V7) at (\x7,\y7) ;
\coordinate (V8) at (\x8,\y8) ;
\coordinate (V9) at (\x9,\y9) ;
\coordinate (V0) at (\x0,\y0) ;

\coordinate (X1) at ($(V1)!0.5!(V4)$); 
\draw[gray,fill](X1)circle(0.08) node[above]{$X_1$};


\draw (V1) -- (V2);
\draw (V1) -- (V3);
\draw (V1) -- (V4);
\draw (V6) -- (V4);
\draw (V5) -- (V4);
\draw (V2) -- (V7);
\draw (V2) -- (V8);
\draw (V3) -- (V9);
\draw (V3) -- (V0);

\draw[red,fill](V1)circle(0.12) node[right]{$Y_1$};
\draw[blue,fill](V2)circle(0.12)  node[left]{$Y_2$};
\draw[olive,fill](V3)circle(0.12) node[left]{$Y_3$};
\draw[violet,fill](V4)circle(0.12) node[left]{$Y_4$};
\draw[teal,fill](V5)circle(0.12) node[left]{$Y_5$};
\draw[purple,fill](V6)circle(0.12) node[left]{$Y_6$};
\draw[magenta,fill](V7)circle(0.12) node[left]{$Y_7$};
\draw[brown,fill](V8)circle(0.12) node[above left]{$Y_8$};
\draw[orange,fill](V9)circle(0.12) node[below left]{$Y_9$};
\draw[cyan,fill](V0)circle(0.12) node[left]{$Y_{10}$};


\end{tikzpicture}
\caption{A subsection of a Voronoi street system $S$ with ten relays $Y_1,\dots,Y_{10}$ and a single user $X_1$. A relay occupies a crossroads with probability $p$, while users are located on streets to according to homogeneous Poisson point process with (linear) intensity $\lambda$. In this example the street system (or graph) has relays $Y_1,\dots,Y_{10}$ at all its  crossroads (or vertices). When we incorporate interference into the connectivity model, we see that adding a single user $X_1$ to the street (or edge) $e_{1,4}$ 
may open the street, if it had been previously closed, in terms of multi-hop communication. But then the same user $X_1$ will also increase the interference at crossroads $Y_1$ and $Y_4$, which may or may not result in closing some or all of the other streets $e_{1,3}$, $e_{1,2}$, $e_{4,5}$ and $e_{4,6}$. But the connectivity of the other streets
will  not be affected by user $X_1$.}
\label{fig.street}
\end{figure}

We assume a homogeneous Poisson point process $X_S$ with (spatial) intensity $\lambda_S>0$ existing in the plane $\R^2$. We interpret the Voronoi tessellation $S$ constructed from this Poisson point process as a street system (or layout). For our Poisson-Voronoi street system $S$, we write $E\eqdef  (e_i)_{i\geq 1}$ and $V\eqdef  (v_i)_{i\geq 1}$ to denote the respective sets of edges and vertices. Of course, in everyday language, the edge-set $E$ and the vertex-set $V$ are respectively the streets and crossroads (or street intersections) in our street system $S$. We have illustrated a part of this model in Figure~\ref{fig.street}. 

\subsubsection*{Users $\User$ and Relays $Y$}
We scatter users (with a single operator) on each street according to a one-dimensional Poisson point process. More specifically, for a given street system $S$ with edges $E$, on each edge $e\in E$ there exists an independent homogeneous Point process of users with  (linear) intensity $\lambda$. The (fixed) relays (belonging to a single operator) are randomly placed at the crossroads independently with probability $p$. More precisely, for the (fixed) relays, we introduce a (doubly stochastic) Bernoulli point process $Y$ on the crossroads $V$ with parameter $p$. We assume the users $\User$ and relays $Y$ are conditionally independent. We write $Z\eqdef  \User\cup Y$ to refer to the superposition of users and relays. It is this point process for which we will study percolation under our proposed connectivity (or communication) model. 

%
%
%

\subsection{Connectivity model}
Under our connectivity model, two distinct points (users or relays)  $Z_i, Z_j\in Z$, where $i\neq j$, are connected provided the following two conditions.
\textbf{Line-of-sight:} Points  $Z_i$ and $Z_j$ are on the same street, meaning there exists an edge $e\in E$ such that $Z_i\in e$ and $Z_j\in e$. 
\textbf{Sufficient signal-to-interference:} The respective signal-to-interference  values of the two points $Z_i$ and $Z_j$ are larger than some threshold $\tau$, meaning for $Z_i\neq Z_j$, the condition
\begin{equation}
\SINR(Z_i,Z_j) \geq \tau\,  \mbox{  and  } \SINR(Z_j,Z_i) \geq \tau  \,,
\end{equation}
where 
\begin{equation}
    \SINR(Z_i,Z_j) = \frac{P_{i,j}} {N+\theta I_{j,i}  }\,,
\end{equation}
and the interference term 
\begin{equation}
 I_{j,i}=\sum_{\mathclap{Z_k\in Z|_e\setminus\{Z_j\} }} P_{k,j} -P_{i,j}\,,
\end{equation}
and similarly for $ \SINR(Z_i,Z_j) $. Here $Z|_e=Z\cap e$ denotes the  users and relays $Z$ located only on street $e$.

\cyan{Our connectivity model will generate a graph $\mathcal{G}$ of streets and crossroads, which is clearly a subgraph of the Poisson-Voronoi street system $S$, meaning $\mathcal{G}\subset S$. 
In the zero-interference case $\theta=0$, our model reduces to the Gilbert-type Poisson-Voronoi model proposed by Le Gall et al.~\cite{le2021continuum}, where the Gilbert radius $r=\ell^{-1}(\noise\tau/P)\,$. Clearly its connectivity graph $\mathcal{G}_r$ is  a subset of $\mathcal{G}$, meaning the Gilbert-type Poisson-Voronoi model is more \emph{percolatable}, it  percolates more easily than our model. }


\subsection{Model considerations}
\subsubsection{Non-monotonic connectivity}\label{sec.nonmonotonic}
What makes a signal-to-interference model challenging in general is its inherent non-monotonic nature. For example, given some receiver, increasing the signal strength of a single transmitter will increase the value of the signal-to-interference in relation to that transmitter, but it will decrease the corresponding values of all the other transmitters. In other words, if a signal-to-interference-based network is altered to create a new connection, other connections may be broken in the process. Creating a new connection in the network may not increase the overall connectivity, whereas in  more purely geometric models new connections generally increase the overall network connectivity.

\subsubsection{Adding relays and users}
Another important aspect of our model is the effects due to the interference at crossroads coming from adjacent streets. In terms of connectivity, adding a relay can open a crossroads, thus potentially opening a previously closed street. However, even if a street becomes open with this newly added relay, the new relay  can still create too much interference, closing previously open streets. Adding a relay can open streets, close streets, or have no effect on street connectivity.

There is a noteworthy distinction between placing additional relays and users in a street system. Adding a user to a street $e$ can either open or close street $e$, or have no effect. But the signal from the newly added user will increase interference at the crossroads at both ends of the street $e$, and these street ends are crossroads; see Figure~\ref{fig.street}. Adding a user to street $e$ can open or close street $e$, but it can only close the immediately adjacent streets of street $e$, in addition to having no effect on street connectivity. 

\subsubsection{Poisson-Voronoi statistics}
Researchers have derived analytic expressions for many of the statistics of the Voronoi tessellation (or mosaic) based on a homogeneous Poisson point process. For a Poisson-Voronoi tessellation on the plane $\R^2$, the \emph{intensity} (or the average density) of vertices is $
\lambda_{0} = 2 \lambda_S$, where $\lambda_S$ is the (spatial) intensity of the underlying Poisson point process. 

When we consider a single Voronoi cell, we speak of the typical cell, where typical is made formal by the concept of the Palm distribution in point process theory. 
The average length of the typical edge in the Poisson-Voronoi tessellation is $
\ell_1=2/(3\sqrt{\lambda_S})\,$. 
The line intensity is the average total edge length per unit area,  given by
$\gamma=2\sqrt{\lambda_S}\,$,
which implies $\ell_1=4/(3\gamma)\,$.
A more mathematically precise definition of $\gamma$ is given by Le Gall et al.\cite[Section II.C.1]{le2019influence}. We also  refer to the standard texts in stochastic geometry by Schneider and Weil~\cite[pages 461 and 477 ]{schneider2008stochastic} or Chiu, Stoyan, Kendall, and Mecke~\cite[pages 357, 368 and 377]{chiu2013stochastic}.

\subsubsection{Scaling and model parameters}
Our street system $S$ coupled with the users  and relays  exhibits scale invariance. 
This invariance inspired  researchers~\cite{le2019influence,le2021continuum} to introduce the dimensionless parameter 
\begin{equation}
    U=\frac{4}{3}\frac{\lambda}{\gamma}=\lambda\ell_1 \,,
\end{equation}
which is the average number of users on a typical street in our street system $S$. In the zero-interference case $\theta=0$, another dimensional parameter appears
 $H=(4/3)/(r\gamma)=\ell_1/r$, which is  the average number of hops to traverse a typical street with length $\ell_1$ under the Gilbert-type model with radius $r$. 

\subsubsection{Losing scale invariance}
The noise term in the signal-to-interference-plus-noise expression~\eqref{eq.def.sinr} removes scale invariance in our model, as the noise is being compared to the signal attenuation through addition. (The noise $\noise$ term can be calculated on different length scales because energy or power scales according to length squared. ) The scale invariance also disappears through the path loss function. Although much research has used the singular and scale invariant path loss model $\ell(d)=1/(d)^{-\beta}$, we need the emitting  power to equal the received power at $d=0$, meaning $\ell(0)=1$ (and we cannot enforce a lower bound on possible distances $d$ between receivers and transmitters). We use a non-singular path loss model such as $\ell(d)=1/(1+d)^{-\beta}$, which is not scale invariant as the distance $d$ is now being added to one. 

\section{Simulation methods}\label{sec.simulation}
Our model has several components depending on different parameters.   The {street system} depends on the parameter $\lambda_S$. The {user locations} depend on the parameters $\lambda_S$ and $\lambda$. The {relay locations} depend on the parameters $\lambda_S$ and $p$. The signal-to-interference  connectivity depends on the path-loss model $\ell$ (which   contains parameters such as $\beta$ and $\scale$) and the parameters $\theta$ and $\tau$, and the noise term $\noise$.  Percolation occurring depends on all these model components and their aforementioned parameters. These observations leads to our simulation method consisting of several steps.

\subsection{Generating a Poisson-Voronoi tessellation}
There are simulations methods for generating Poisson-Voronoi cells without edge effects. But in our percolation model, we want to examine truncated Voronoi tessellations, as we want to study connectivity across a square simulation window. Using such window crossings is one of the standard approaches for studying percolation via simulations~\cite{newman2001fast}.

For our square simulation window $W_1$, we surround it with eight translated windows $W_2,\dots,W_9$ of equal size, creating one large square simulation window $W=\cup_{i=1}^9 W_i$ . We generate a Poisson point process of density $\lambda_S$ on this simulation window $W$. We find the corresponding Voronoi diagram using a standard library function such as \emph{voronoin} in MATLAB or \emph{scipy.spatial.Voronoi} in Python, noting that both use the same underlying algorithm. We then truncate the streets, leaving only streets completely inside $W_1$ or truncated streets that intersect with $W_1$ in an effort to reduce edge effects. (Interference contributions may come from relays and users located outside of $W_1$, while such edge effects are not an issue in the zero-interference Gilbert model.) 

\subsection{Positioning relays and users}
This is the most straightforward step. Every crossroads (or vertex) is a potential relay with probability $p$, so we simple simulate Bernoulli variable (that is, perform a coin toss) for each crossroads. The users are positioned by simply simulating a one-dimensional homogeneous Poisson point process on every street. Users can be located on  streets that extend beyond the simulation window $W$. 

\subsection{Finding open and closed streets }
To find which streets are open and closed, we first need to calculate the signal-to-interference-plus-noise ratio at all relays, where the interference comes from both relays and users on common streets. Checking direct relay-to-relay communication is then straightforward. 

If direct relay-to-relay communication is not possible, then we need to check if communication is possible through users. But for users on any street, there are combination of ways data can be relayed from one street end (or transmitting relay) to the other street end (or receiving relay). The number of combinations soon becomes unwieldy. 

But we can use the fact that if data \emph{cannot} be transmitted from one relay to another relay on a shared street by transmitting through every user on the street (using the maximum number of hops), then it is impossible for the data to be relayed between the two relays by transmitting through some subset of users (using fewer hops). This claim is only true provided, crucially, the path loss model is monotonic. (This claim is no longer true, of course, if we were to include random signal fading.) The reasoning is that, because  all transmitters have the same power, if a signal coming from an immediate neighbour has an insufficiently strong signal-to-interference, then another signal coming in the same direction from a more distant neighbour  cannot have a stronger signal-to-interference.


\subsection{Tracking connectivity of the street system}
We recall that the connectivity in our signal-to-interference model is not monotonic. Other percolation models, such as the Gilbert model, are monotonic, which allows the use of union-find methods to quickly keep track of (connected) components. In lieu of this, we can simply use in-built functions, such as  \emph{conncomp} in MATLAB or \emph{connected\_components} in Python (NetworkX library), to find the largest component of the network, which fortunately can be done quickly on a standard desktop machine.

\subsection{Determining if percolation occurs}
There are a few ways to estimate when percolation occurs in simulations. Researchers detailed three of them in a related paper on device-to-device networks by Garfur et al.~\cite{Gafur2018}. For our work, we will use the popular crossing method, as detailed in the paper by Newman and Ziff~\cite{newman2001fast}. (One of the original appeals of this method is that for a specific percolation problem, closed-form solutions exist for the so-called winding probabilities on a torus.) We assume that percolation occurs when there is a vertical or horizontal crossing (or both) between one side of the square simulation window to another. 

More precisely, we find the streets that intersect with the four sides of the square simulation window (say, north, east, south, and west). Then for, say, a vertical crossing, we simply check if both the north and south sides intersect with streets that  form part of the (same) connected component, and similarly for a horizontal crossing. If so,  we then say percolation has occurred in that simulation, as it is possible to cross from north to south or east to west (or both).  

For each simulation, we note if any crossings occur, and then  take the average to obtain  the percolation probability. If this \emph{connected probability} is close to zero (for some tolerance), we say no percolation occurs for given the parameter regime. Conversely, if the connected probability is close to one (for some tolerance), we say percolation has occurred. As we expand the simulation window, the (empirical) connected probabilities will approach zero and one, respectively.

\subsection{Finding critical values of system parameters}
Using the above approach, we can estimate  when percolation occurs for certain parameter regimes. We simply vary a system parameter and observe when percolation occurs in the network model. Mathematically, when percolation occurs, there will be a sharp change from zero to one (or one to zero) in the phase transition curve, formed from the parameter values and connection probabilities, resembling a step function. 

But this is not the case for simulations, where the phase transition curve resembles a sigmoid curve (from zero to one), such as that of the logistic function, or a reverse sigmoid curve  (from one to zero). We use this fact to estimate critical values of the parameter of some statistic $f(\mu)$ (the percolation probability in our case) that is some function of a system parameter $\mu$. We assume that the sigmoid curve of $f(\mu)$ allows us to fit a statistical model based on the logistic function,  whose inverse is the logit function $h(t)\eqdef\log[1/(1-t)]$, yielding the statistical fitting problem
$   h[f(\mu)]= a\mu +b\,$.
 We fit this logisitic model to the phase transition curve (using standard fitting methods). We then say that the estimated critical value $\mu^*$ occurs at the inflection point of the fitted curve (when the second derivative is zero), which happens when $t=0.5$, meaning $h''(0.5)=0$. In terms of the fitted model's parameters $a$ and $b$, the critical value is estimated as $\mu^*=-b/a$ 

\subsection{Simulation code}
The simulation code is online~\cite{keeler2023voronoicode}. All simulations were performed on a desktop computer, often using multi-core processing. Simulation times ranged from a few minutes to a couple of hours.

\section{Results}
We ran various simulations of our signal-to-interference-plus-noise model based on a Poisson-Voronoi street system. Naturally, when we set $\theta = 0$, giving the zero-interference case, we can completely reproduce the results of the Gilbert-type model with the line-of-sight assumption, such as those given by Gafur et al.~\cite[Table I]{Nila2} and Le Gall et al.~\cite{le2021continuum} for the Poisson-Voronoi street system.

\subsection{Parameter values}
For our simulation results, we seek to use parameter values that reflect those used in models of real-world networks.  We  choose the street intensity $\gamma=2\sqrt{\lambda_S}$ by setting the length of a typical street $\ell_1=4/(3\gamma)=2/(3\sqrt{\lambda_S})$. 
We set the mean street length to be $\ell_1 = 100$ metres, which corresponds to the city center of a medium sized city. We set the signal-to-interference-plus-noise threshold $\tau = 1$, giving a threshold of zero decibels. We set the noise $N = 10^{-8} $ milliwatts  (or about $-81$ dBm) and the signal power $P = 1 $ milliwatt, giving the dimensionless noise term $\bar{N} = 10^{-8}$.  We place a relay at each crossroads by setting $p=1$. We use the interference reduction factor $\theta = 0.004$.

\cyan{We use the path loss function  $\ell(d) = 1/(1+\scale d)^{\beta}$, which was also used by Dousse, Baccelli and Thiran~\cite{dousse2005impact}. For the parameters $\beta$ and $\scale$, we can use the free-space path loss model, due to the  line-of-sight  assumption, giving the values $\beta=2$ and $\scale=(4\pi/\lambda_f)^2$, where $\lambda_f$ is the signal wavelength. For fifth-generation (5G) networks, the scaling parameter $\scale$ will range in the vicinity of one to two hundred inverse metres. For the zero-interference case ($\theta=0$), our model becomes the Gilbert-type model, which can be interpreted as a signal-to-noise model such that $\ell(r)=\bar{\noise}\tau$.  If we set the Gilbert radius $r=\ell_1$, then typically  some users are needed to relay data between two adjacent relays. (The average number of hops will be $H=1$.) Given  $\bar{N} = 10^{-8}$ and $\ell_1=100$ metres, the scaling distance $\scale$ is essentially one hundred inverse metres, because $\ell(100)=1/(1+100\scale)^2= \bar{\noise}\tau=10^{-8}$, so we set $\scale=100 $ inverse metres.}

\subsection{Numerical results}

\begin{figure}
    \centering
    \includegraphics[scale = 0.22]{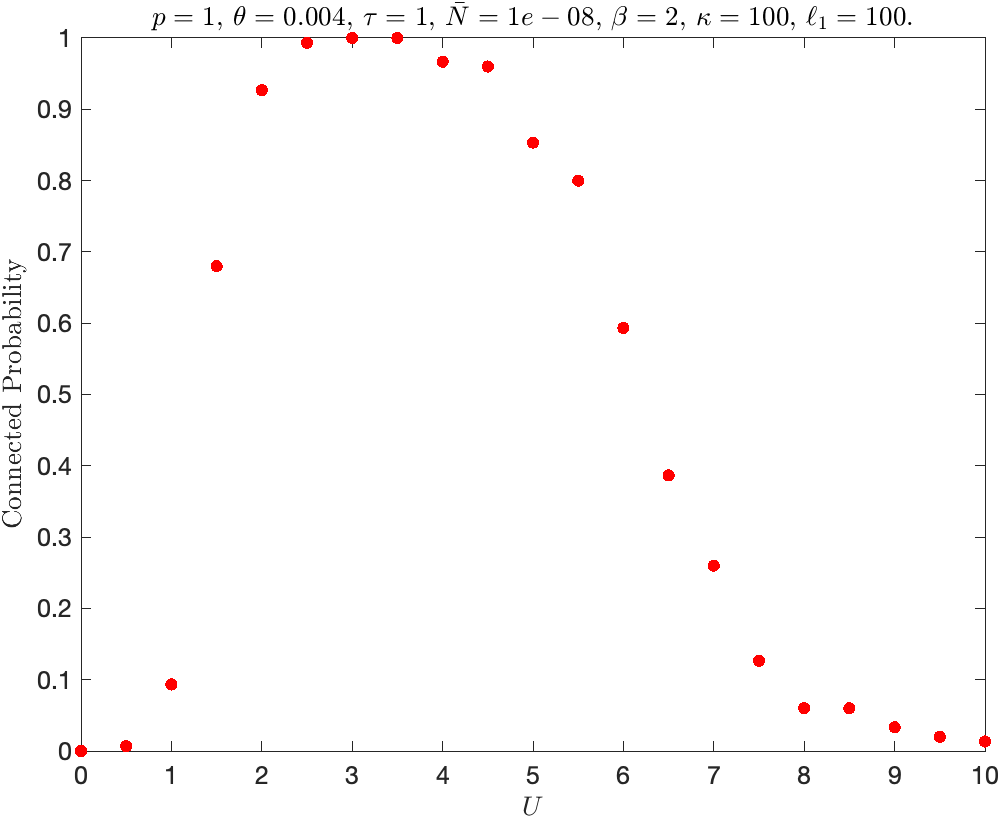}
    \caption{$p = 1$ and $\theta = 0.004$. The double criticality phenomenon appears clearly here. The plateau between the two critical points is not very large : percolation disappears due to interference for a moderate intensity of users}
    \label{fig:basic}
\end{figure}

Our parameter values gives us the curve in Figure~\ref{fig:basic}, showing connected probabilities plotted as a function of the user numbers $U$ on a typical street. There is no percolation when there are no users, as the adjacent relays are too far apart to transmit data. Adding users gives us our first (upward) sigmoid curve, showing a sudden jump in connectivity as percolation occurs in the network model. The critical value $U_1^*$ for this percolation is about two users on a typical street of length one hundred metres. This curve features a plateau corresponding to a percolation regime. After percolation occurs, a second (downward) sigmoid appears, indicating the sudden loss of percolation (and hence connectivity) due to interference from users. This happens when the  critical value $U_2^*\approx 6$, meaning there are about six users (with the same operator) on a typical street. 

\begin{figure}
    \centering
    \includegraphics[width=.22\textwidth]{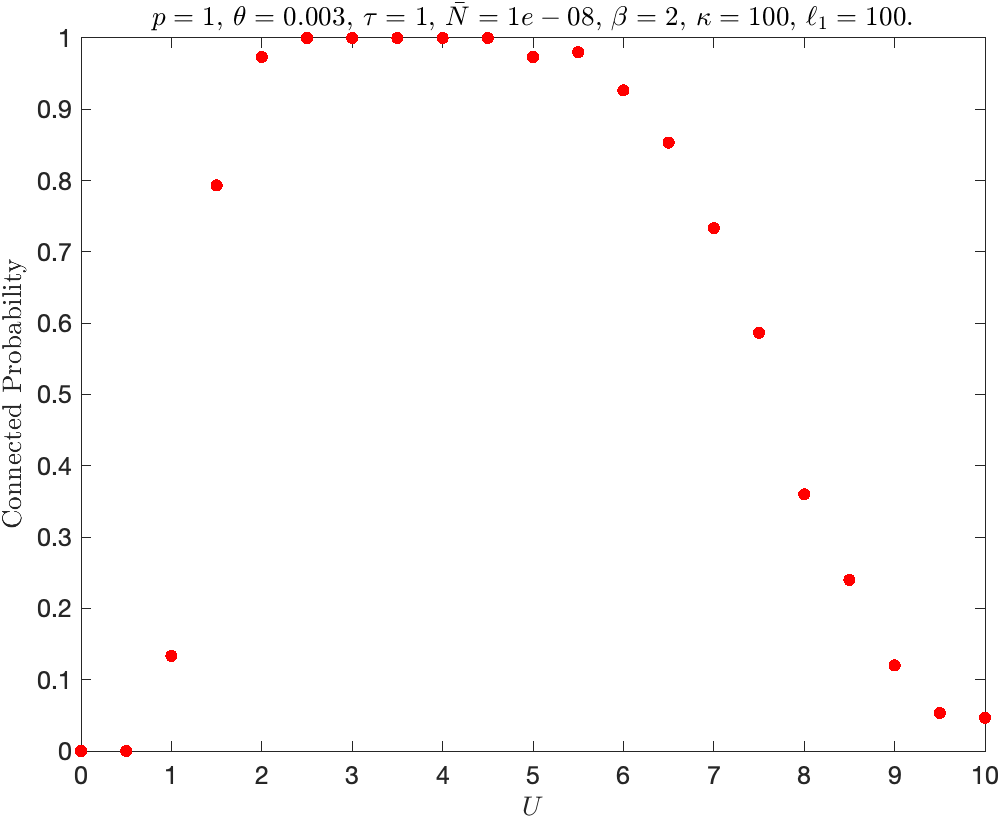} 
    \includegraphics[width=.22\textwidth]{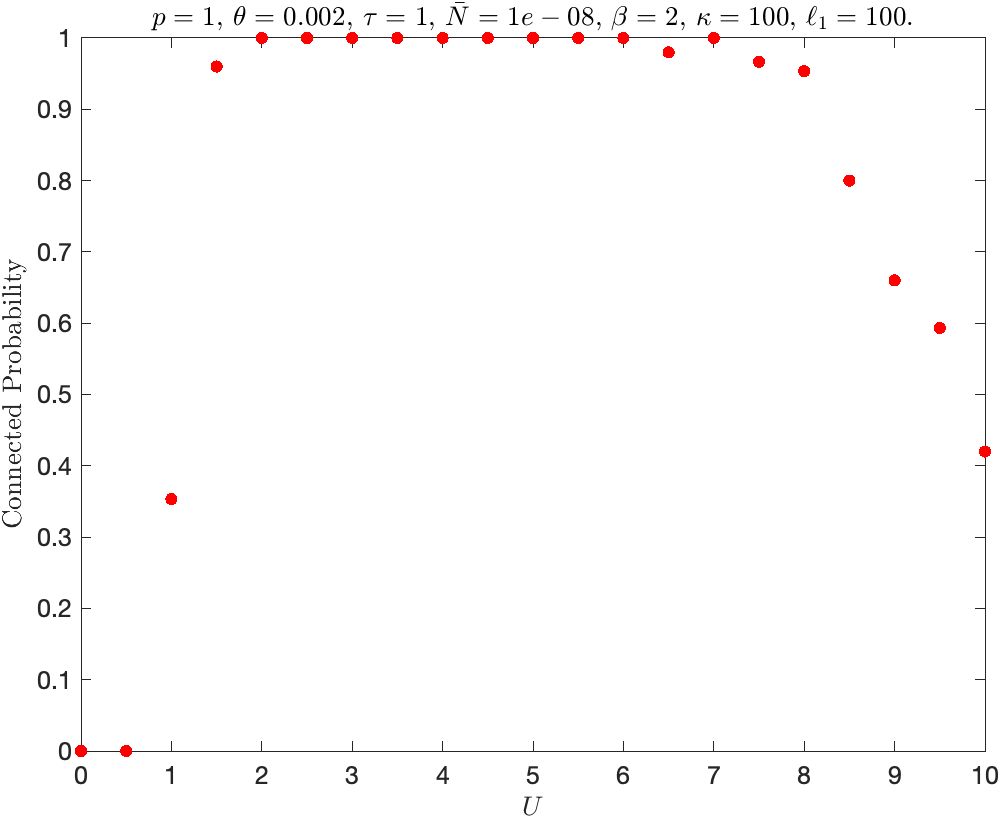} 
    \caption{Decreasing interference reduction factor $\theta$  from  $\theta=0.003$ (on the left) to $\theta=0.002$ (on the right), we see that the plateau enlarges, due to the  reduction in interference.}
    \label{fig:change_theta}
\end{figure}

In Figure~\ref{fig:change_theta} we see that decreasing $\theta$ decreases the interference. On the one hand, this slightly helps percolation to happen (indicated by $U_1^*$ decreasing). But on the other hand delays,  decreasing $\theta$  also delays the disappearance of percolation, as the value of $U_2^*$ increases. Hence, decreasing $\theta$  enlarges the plateau the on both sides.

\begin{figure}
    \centering
    \includegraphics[width=.22\textwidth]{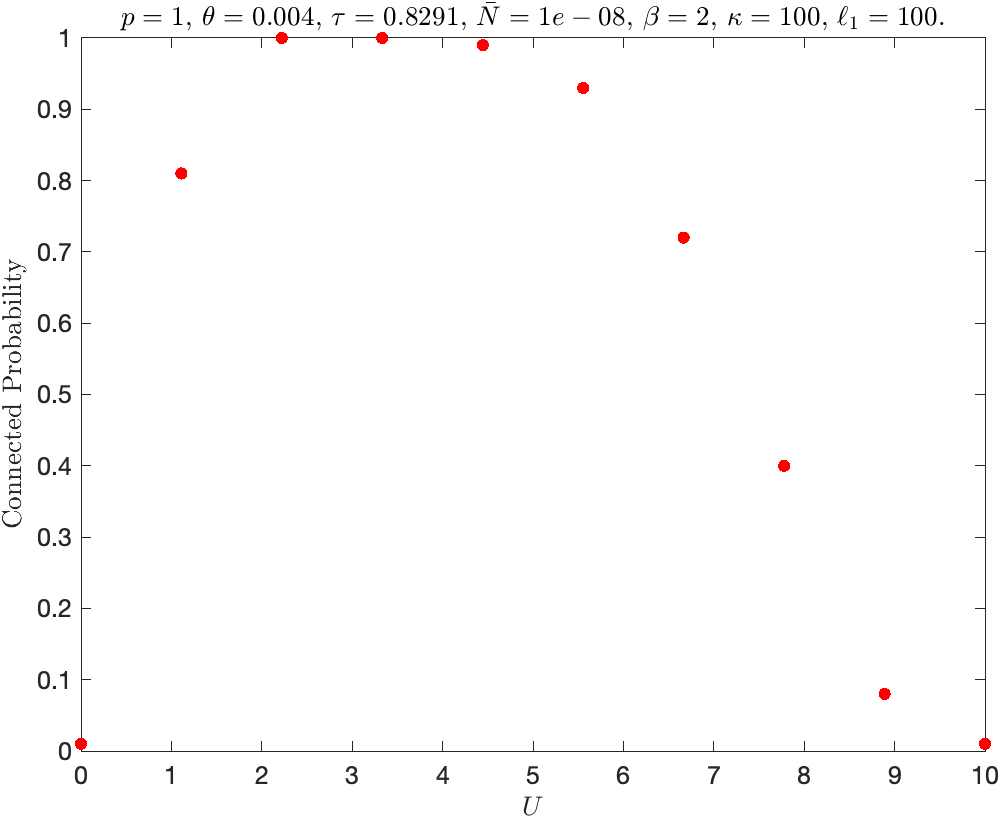}
    \includegraphics[width=.22\textwidth]{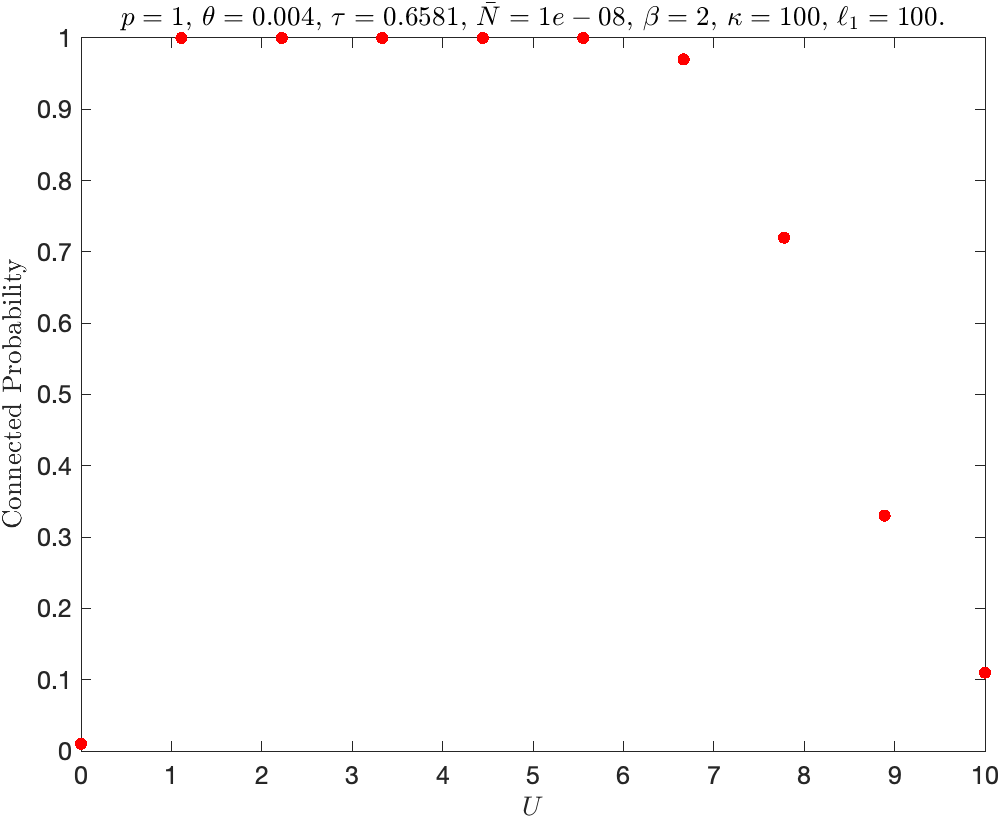}
    \caption{Decreasing the threshold $\tau$ from  $\tau=0.829$ (on the left) to  $\tau=0.658$ (on the right), we see that the plateau grows in size because is now a weaker requirement on the signal-to-interference-plus-noise.}
    \label{fig:change_tau}
\end{figure}

 We also observe the effect of varying the the signal-to-interference-plus-noise threshold $\tau$. In Figure~\ref{fig:change_tau}, we see that decreasing the threshold $\tau$ allows for connections with weaker signals, which helps the network model percolate. We can compare these results with those in Figure~\ref{fig:basic}, indicating that $U_1^*$ decreases and $U_2^*$ increases as the threshold $\tau$ decreases.

\begin{figure}
    \centering
    \includegraphics[width=.22\textwidth]{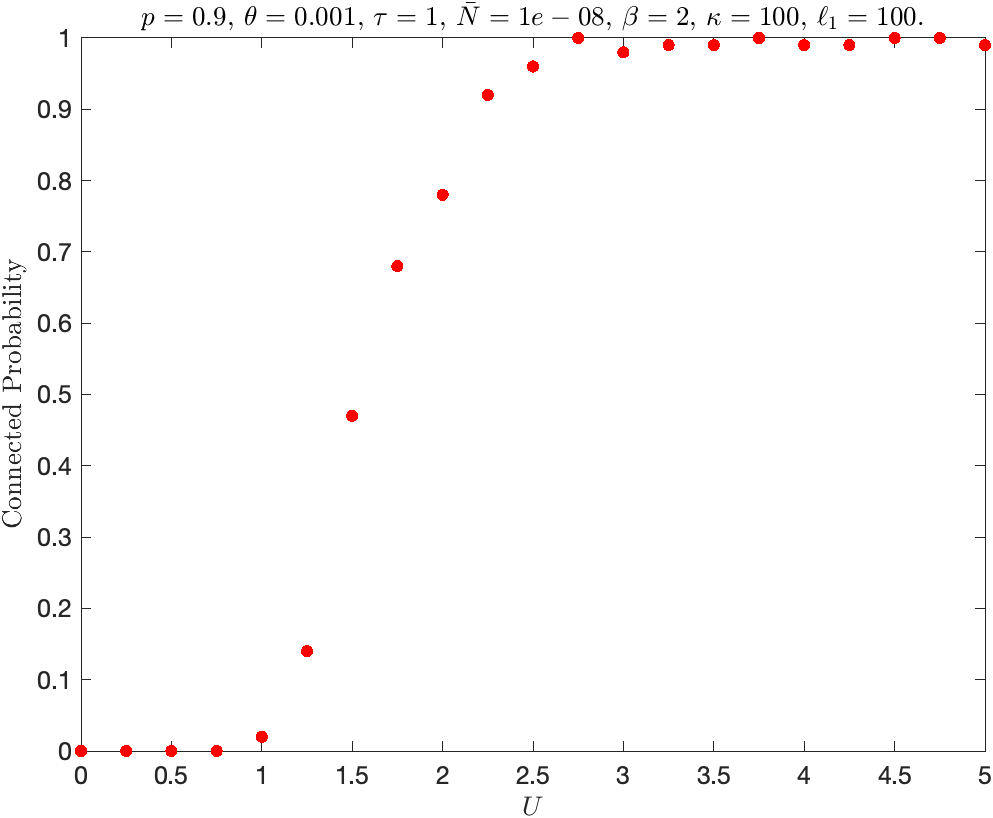}
    \includegraphics[width=.22\textwidth]{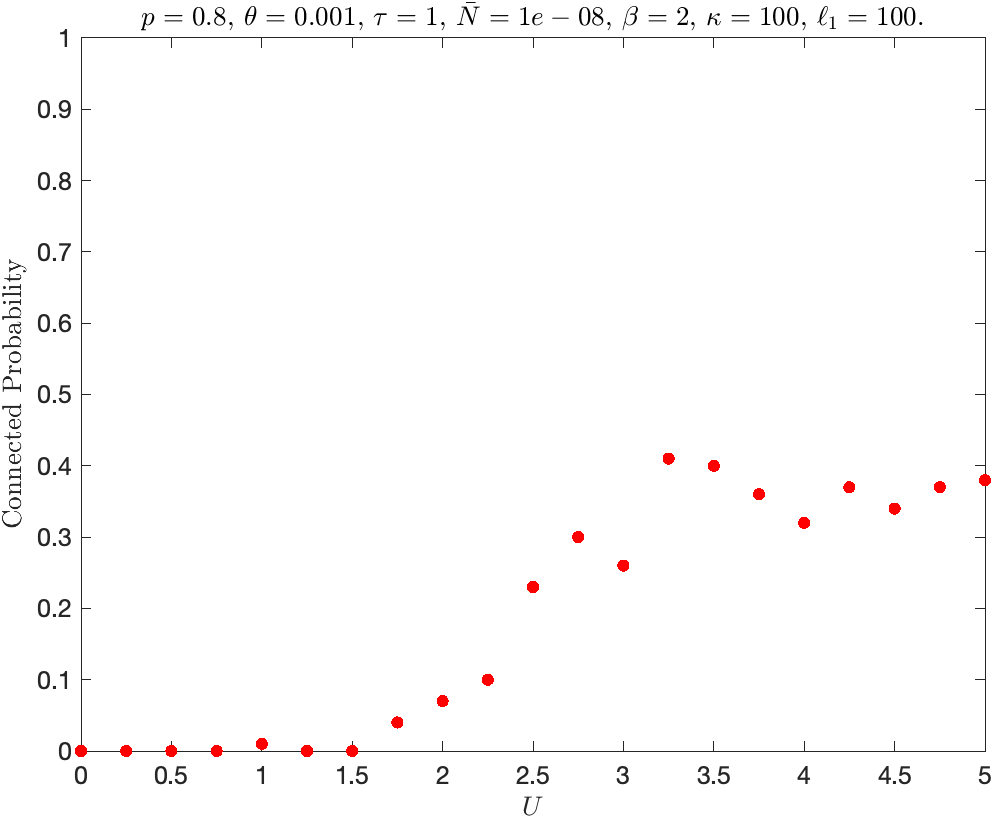}
    \caption{ As we decrease relay probability  $p$ from $p = 0.9$ (on the left) to $p = 0.8$ (on the right),  percolation in the network almost disappears, as there are fewer relays.}
    \label{fig:change_p}
\end{figure}

\begin{figure}
    \centering
    \includegraphics[width=.22\textwidth]{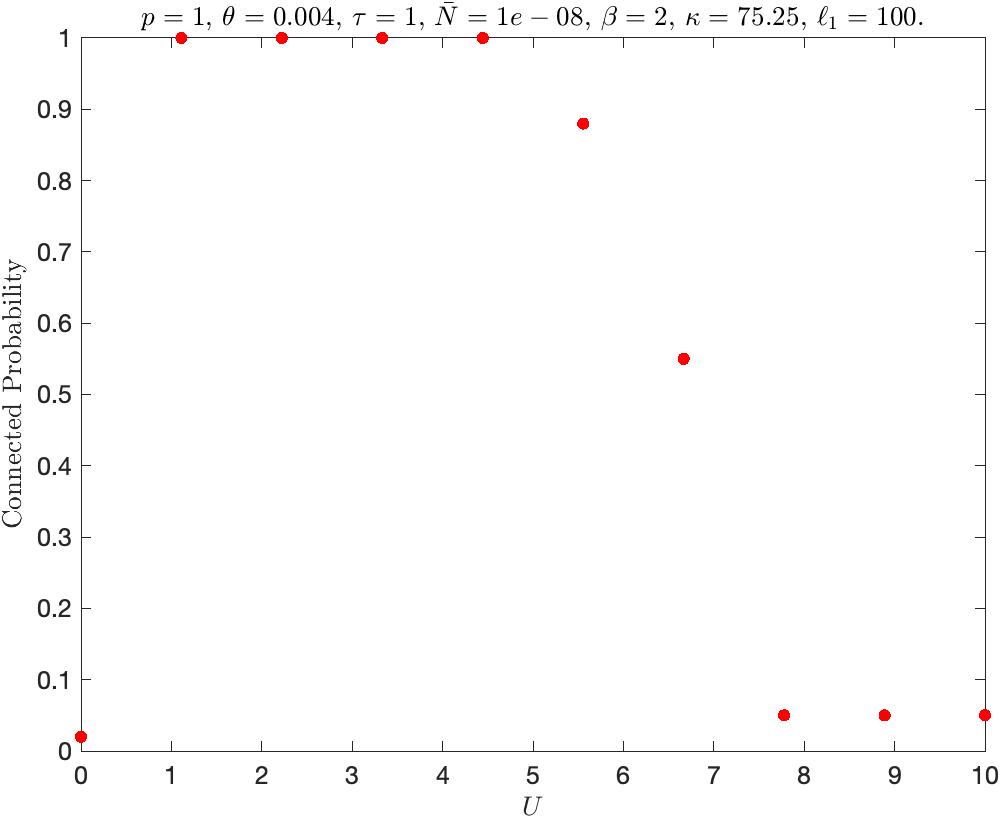}
    \includegraphics[width=.22\textwidth]{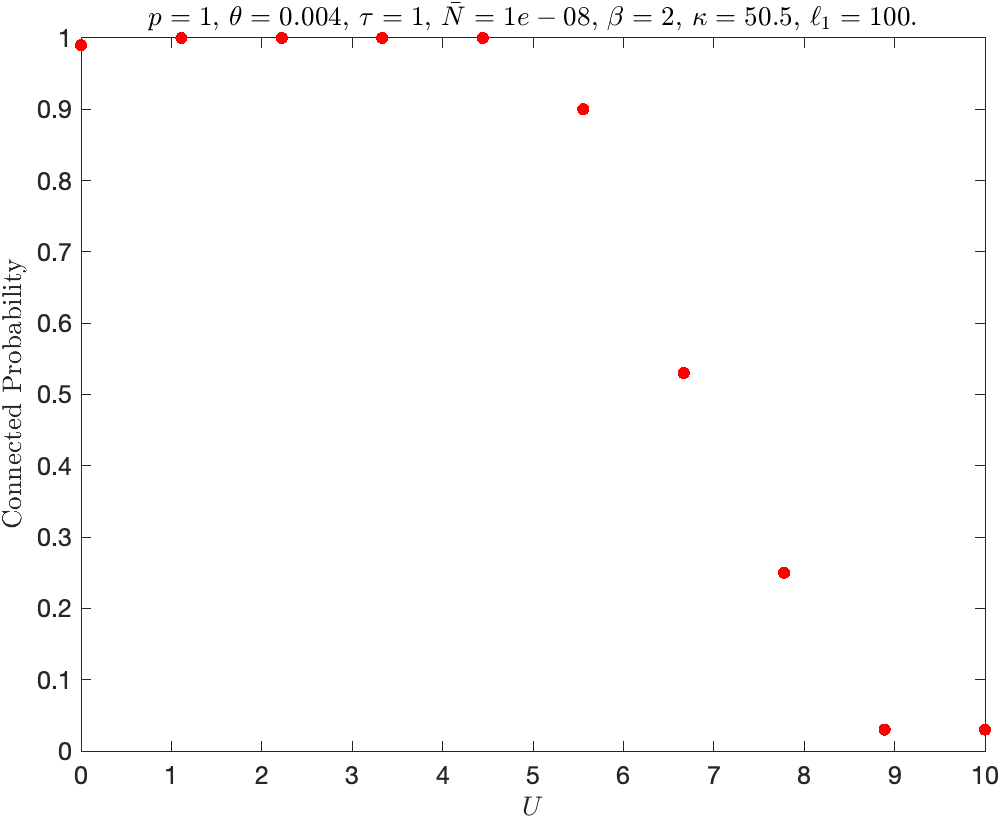}
    \caption{Decreasing the scaling parameter $\scale$ from $\scale=75.25$ (on the left) to  $\scale=50.50$ (on the right) helps percolation to occur.}
    \label{fig:change_K}
\end{figure}

We also observe other effects from varying the model parameters. Figure~\ref{fig:change_p} shows decreasing $p$ slows the appearance of percolation, as there are fewer relays. 

Figure~\ref{fig:change_K} shows that increasing the (path loss) scaling parameter $\scale$ helps percolation. Indeed, decreasing $\scale$ increases the Gilbert radius $r_0$, so that percolation becomes easier, compared to the percolation curve Figure~\ref{fig:basic}. As we increase $\scale$, see that $U_1^*$ decreases and $U_2^*$ increases.

\begin{figure}
    \centering
    \includegraphics[width=.22\textwidth]{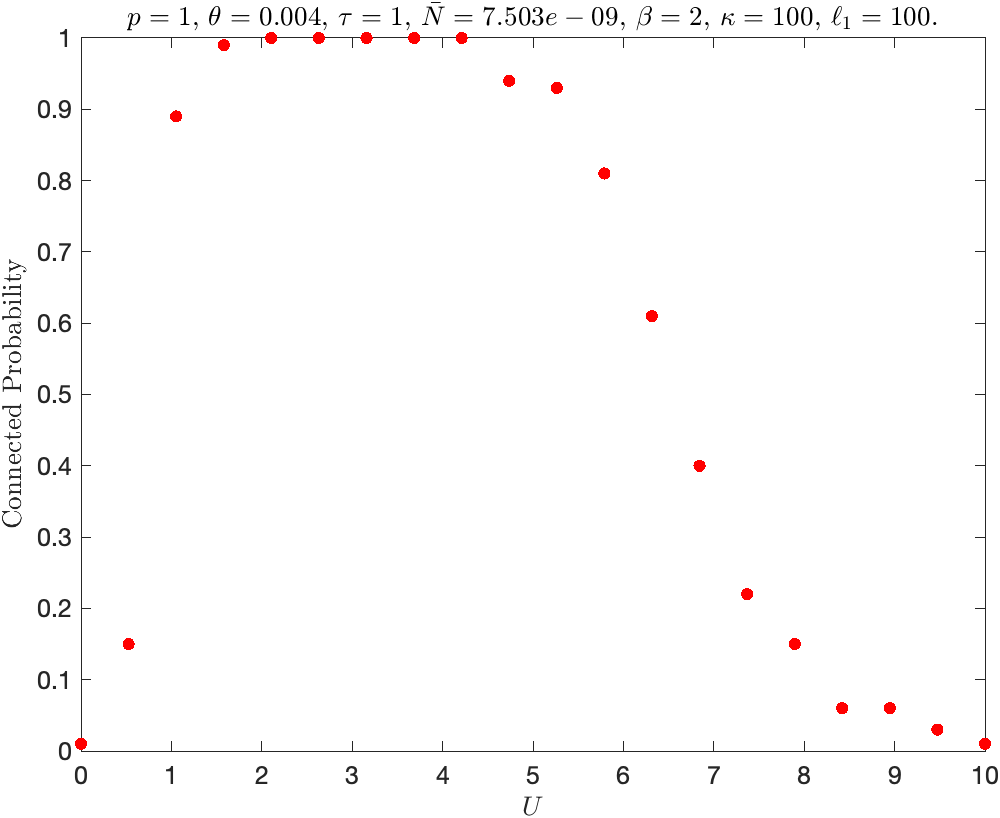}
    \includegraphics[width=.22\textwidth]{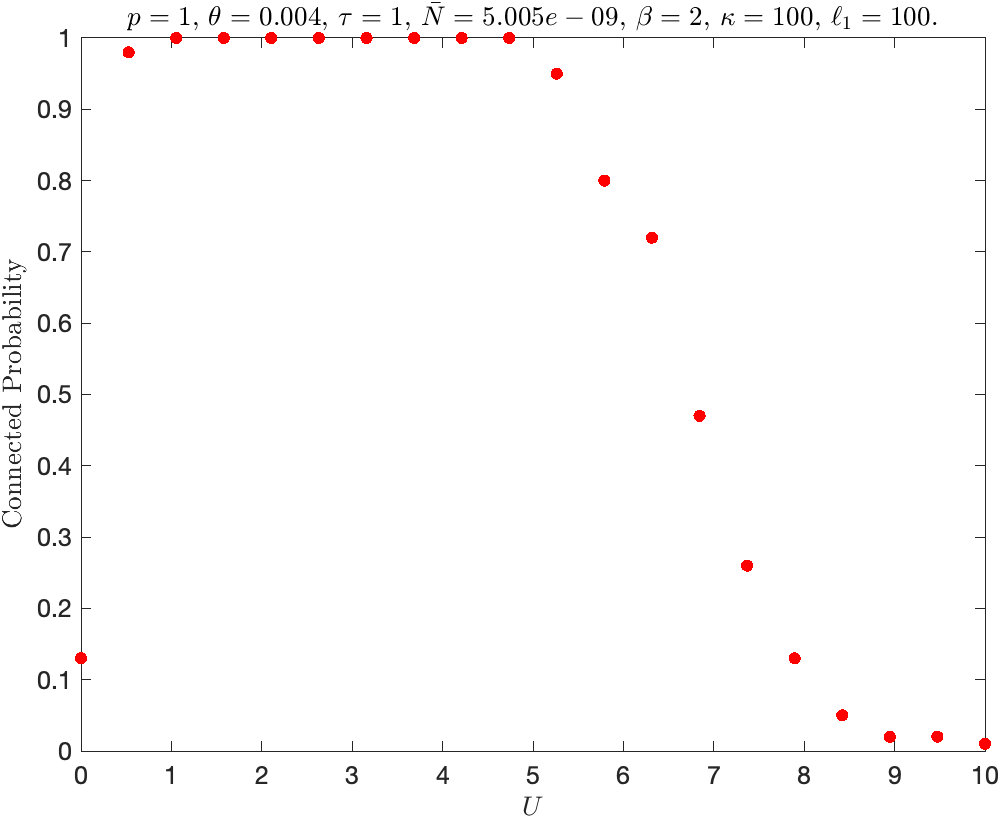}
    \caption{Decreasing the noise ratio $\bar{N}$ from $\bar{N} = 0.75 10^{-8}$ (on the left) to  $\bar{N} = 0.5 10^{-8}$ (on the right).}
    \label{fig:change_Power}
\end{figure}

We see in Figure~\ref{fig:change_Power} that increasing $P$ (by decreasing the dimensionless noise term $\bar{N}$) also helps percolation. Again, compare this percolation curve to that in Figure~\ref{fig:basic}, we see that $U_1^*$ decreases and $U_2^*$ increases.

\begin{figure}
    \centering
    \includegraphics[width=.22\textwidth]{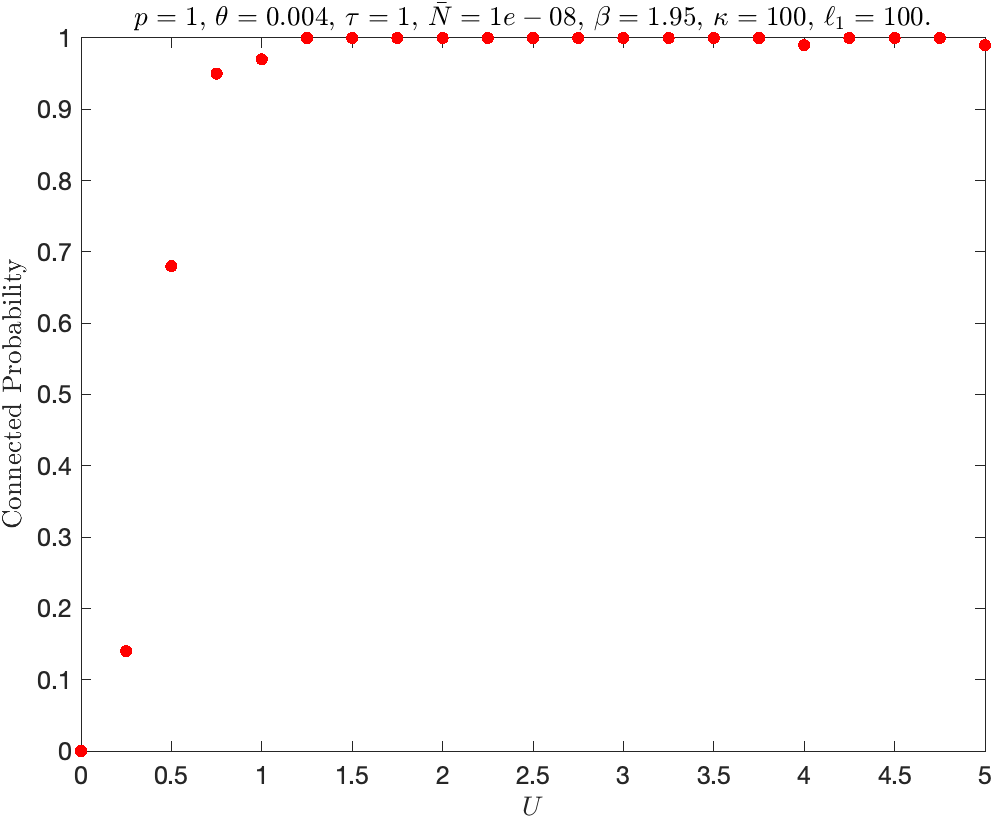}
    \includegraphics[width=.22\textwidth]{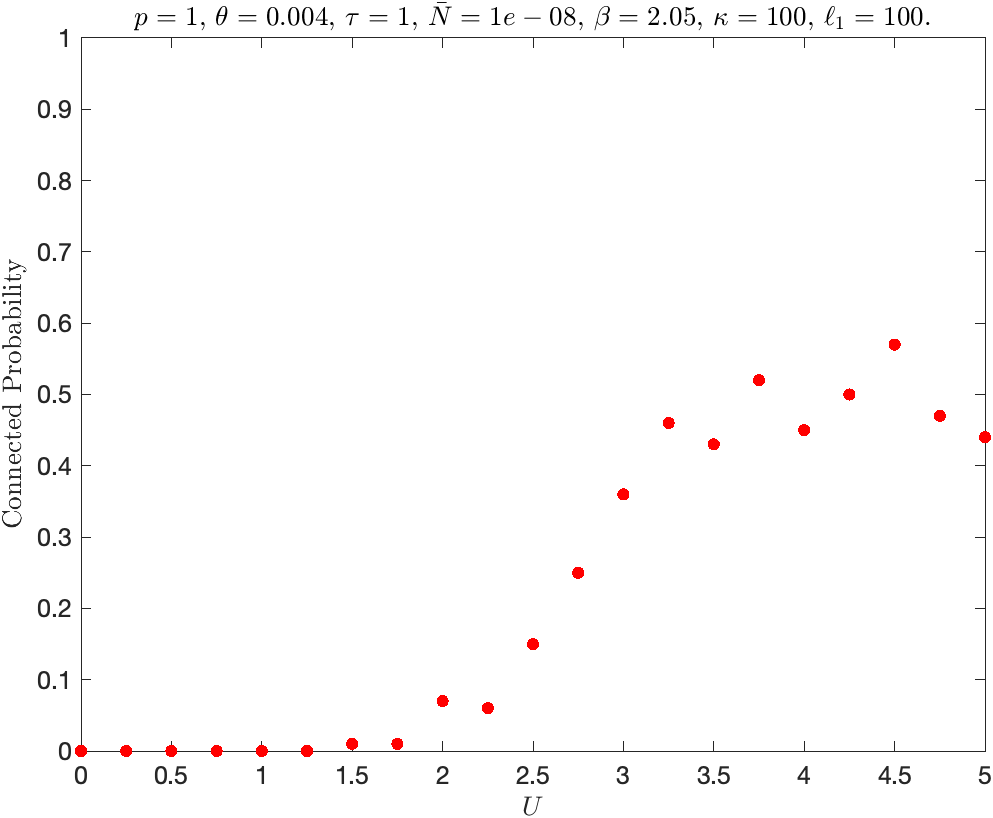}
    \caption{$\beta = 1.95$ and $\beta = 2.05$. Two very different profiles for very close values of $\beta$}
    \label{fig:change_beta}
\end{figure}

In Figure~\ref{fig:change_beta} we observe that very small changes in the path loss exponent $\beta$ have tremendous effects. Increasing $\beta$ means decreasing $\ell(d)$, which means at the same time decreasing the effects of interference and decreasing the received power. But there are fewer interferers present, because $U$ is low, and their distances to a receiver is greater than that of any transmitter, such that the decrease in the received power is the main effect, tending to break percolation as $\beta$ increases. Since $\beta = 2$ seems very near to the $\beta$ threshold for percolation, the parameters should be tuned thoroughly to study a given real situation.

\begin{figure}
    \centering
    \includegraphics[scale = 0.22]{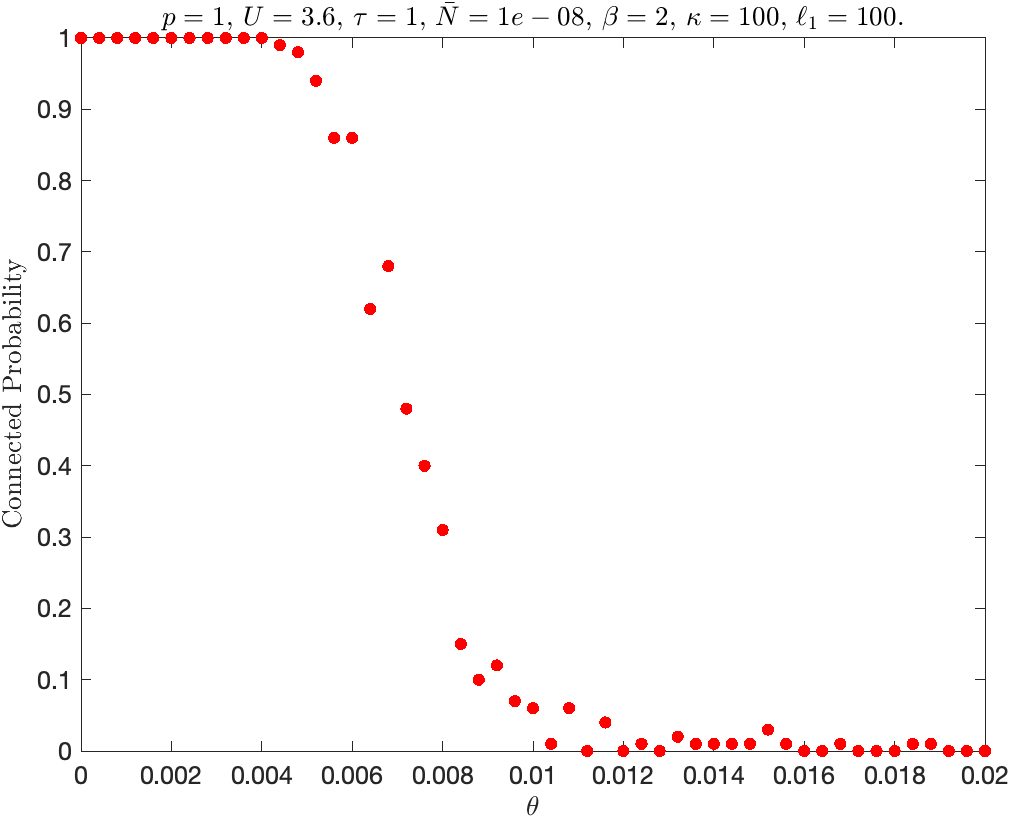}
    \caption{For $U = 3.6$, we can vary $\theta$ to find its critical value.}
    \label{fig:U3.6}
\end{figure}

From another point of view, for fixed $U$, we can look for the $\theta$ values  that allow percolation. We see in Figure~\ref{fig:U3.6} that for $U = 3.6$ (a value generally near the middle of the plateau in the preceding figures), a critical value of $\theta$ appears around $0.007$.

\begin{figure}
    \centering
    \includegraphics[scale = 0.25]{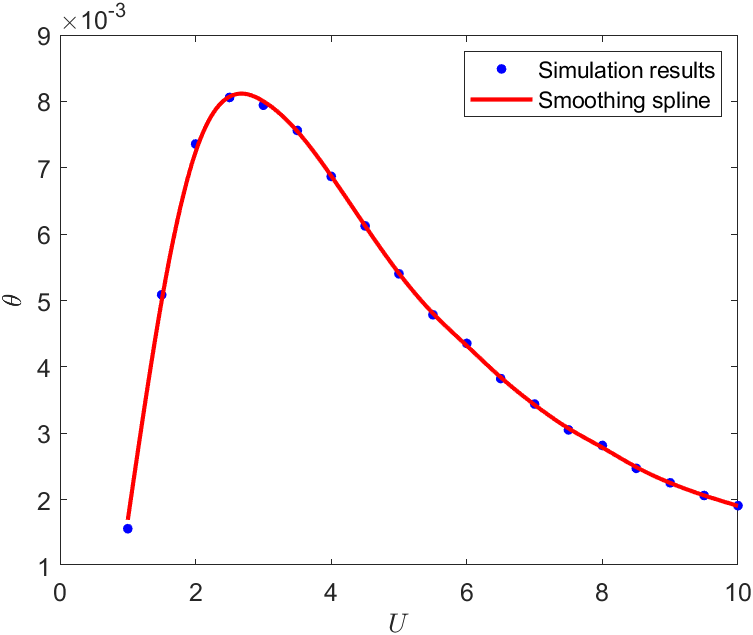}
    \caption{By plotting $\theta$ as a function of $U_1^*$ and $U_2^*$, we create a phase diagram for our percolation model. For our network model, percolation occurs under the curve, whereas percolation does not occur above the curve.}
    \label{fig:thetaU}
\vspace{-1ex}
\end{figure}

Using the above approach of finding $\theta$ values, we can study how the critical thresholds $U_1^*$ and $U_2^*$ evolve as we vary $\theta$, as illustrated in Figure \ref{fig:thetaU}. We see a curve very similar to the one presented by  by Dousse, Baccelli and Thiran ~\cite[Figure 3]{dousse2005impact}. Above the curve is the subcritical regime, where percolation does not occur, whereas under the curve is the supercritical regime, where percolation does occur. But the critical values that make up the critical regime also depend on the parameters $p$, $\beta$ and $\scale$, as we showed above.

\subsection{Pole capacity and a bound on $U_2^*$}
We can derive an upper bound on the second critical value $U_2^*$, which in turn gives an approximate upper bound on the  $\theta(U)$ curve plotted in Figure~\ref{fig:thetaU}. The number of connections received simultaneously on any receiver cannot be greater than $M\eqdef1+1/(\theta\tau)$. Owing to the definition of the signal-to-interference-plus-noise ratio and simple algebra, this simple bound holds true regardless of any other parameter values, including those of the  path loss model. This observation is sometimes called the \emph{pole capacity}; for example, see the monograph~\cite[Lemma~5.1.2]{hetnetbook}, as well as its application in the aforementioned percolation work by Dousse, Baccelli and Thiran~\cite[Theorem 1]{dousse2005impact}.

For our device-to-device network model, this bound implies that a user, say, $x$ can connect up to $M/2$ users behind user $x$  and up to $M/2$ users in front of user $x$, while trying to relay data between some of the users. 
On average, these $M/2$ users give  user $x$ a communication range of $1/\lambda\times M/2$ distance (behind and in front). Users on every street form a Poisson point process with intensity $\lambda$, so we can on average find a user over a distance $1/\lambda$.   When the user density $\lambda$ increases, the communication range  $1/\lambda\times M/2$ decreases to zero. This relationship captures the decrease in communication range due to the increasing interference as the user numbers increase. However, if there are too many users, the communication range becomes so small that even the (more percolatable) Gilbert-type Poisson-Voronoi  model ($\theta = 0$) does not percolate. 

For this Gilbert-type  model, let $r^*=r^*(\lambda)$ denote the critical range for the user density $\lambda$.  We have a necessary condition, such that that $\lambda$ is not too large to make the range too small, a requirement that is roughly expressed above as  $r^*(\lambda) > 1/\lambda\times M/2$.  We can reformulate this condition in terms of the number of users $U$ per typical street  and the number of hops $H$ needed to traverse the typical street with Gilbert radius  $r$.  We write $U^*=U^*(H)$ to denote the critical number of users per typical street needed for percolation in the Gilbert-type Poisson-Voronoi  model, which is a function of $H$. This leads to the bound
\begin{equation}
   U^*(H)<\frac{HM}{2}=\frac{H}{2}(1+1/(\theta\tau)))\,.
\end{equation}
 For small values of the product $\theta\tau$, we obtain the bound $U^*(H)<(\theta\tau)^{-1}/2 H$. The function $U^*(H)$ is plotted in the related work by Le Gall et al.~\cite[\cyan{Figure~5a}]{le2019influence}. This function can be used to derive a bound on the relation between $\theta\tau$ and $U$ for the percolation. Our necessary condition $U^*(H)<H\theta\tau/2 $ is equivalent to  $U^*(H)$ being below the linear function $ H(\theta\tau)^{-1}/2$. For example, for  $\tau\theta=6/1000$ we obtain the linear function $1000/(2\times6)=83$, giving gives a line for $H\theta\tau/2 $ that is too steep to be able to observe it intersecting  with the curve $U^*(H)$ given in the aforementioned work~\cite[\cyan{Figure~5a}]{le2019influence}. The linear function $10/3.5 H=2.8 H$ intersects with $U^*(H)$ at $H=3.5$, predicting the maximum $U^*(3.5)=10$ users per street for percolation with the value $(1+1/\tau\theta)/2=2.8$,  resulting in $\theta\tau=0.217$. This line of inquiry deserves further attention beyond the scope of the current work. 

\section{Conclusion}
We presented a new model by incorporating interference into the line-of-sight (continuum) percolation model presented by Le Gall, B{\l}aszczyszyn, Cali and En-Najjary~\cite{le2021continuum}. Given interference, we observed \blue{
}
the double phenomenon of both percolation appearance and disappearance  in our network model when varying the user density. Users are needed to relay data along longer streets, but too many users decreases connectivity due to increased interference. 

We also showed that percolation depends mainly on the interference reduction  parameter $\theta$ whose critical value depends on the other model parameters. However, the parameters can vary greatly due to, for example, the wide frequency range of fifth-generation networks. We note that network researchers and operators should study the network with the relevant parameters to determine the maximum value of the interference reduction parameter $\theta$ for enabling network percolation and, hence, network connectivity. One natural line of inquiry is to adapt the street models, such as a Poisson-Delaunay tessellation or Poisson line process, for different city types.




\end{document}